\documentstyle[12pt,equation]{article}
\include{margin}
\newcommand{\ww}{\mbox{$W^+W^-$}}
\newcommand{\eplem}{\mbox{$e^+e^-$}}
\newcommand{\mslr}{\mbox{$m_{\tilde{l}_R}$}}
\newcommand{\msll}{\mbox{$m_{\tilde{l}_L}$}}
\newcommand{\wi}{\mbox{$\widetilde{W}_1$}}
\newcommand{\zi}{\mbox{$\widetilde{Z}_1$}}
\newcommand{\zin}{\mbox{$\widetilde{Z}_2$}}
\newcommand{\mwi}{\mbox{$m_{\widetilde{W}_1}$}}
\newcommand{\mzi}{\mbox{$m_{\widetilde{Z}_1}$}}
\newcommand{\mzin}{\mbox{$m_{\widetilde{Z}_2}$}}
\newcommand{\msn}{\mbox{$m_{\tilde{\nu}}$}}
\newcommand{\slrp}{\mbox{$\tilde{l}_R^+$}}
\newcommand{\slrm}{\mbox{$\tilde{l}_R^-$}}
\newcommand{\slr}{\mbox{$\tilde{l}_R$}}
\newcommand{\cb}{\mbox{$\cos \! \beta$}}
\newcommand{\sib}{\mbox{$\sin \! \beta$}}
\newcommand{\cw}{\mbox{$\cos \! \theta_W$}}
\newcommand{\sw}{\mbox{$\sin \! \theta_W$}}
\newcommand{\etmiss}{\mbox{$\not \!\! E_T$}}
\newcommand{\tanb}{\mbox{$\tan \! \beta$}}
\newcommand{\mx}{\mbox{$M_X$}}
\newcommand{\be}{\begin{equation}}
\newcommand{\ee}{\end{equation}}
\newcommand{\een}{\end{subequations}}
\newcommand{\ben}{\begin{subequations}}
\newcommand{\beq}{\begin{eqalignno}}
\newcommand{\eeq}{\end{eqalignno}}
\begin{document}
\pagestyle{empty}
\begin{flushright}
MAD--PH--881 \\
DO--TH 94/28 \\
March 1995
\end{flushright}
\vspace*{0.25truein}

\begin{center}
{\Large\bf HUNTING VIRTUAL LSPs AT LEP 200
 \\[0.5truein]}
{\large Amitava Datta\footnote{Present address: Inst. for Physics, Univ
Dortmund, D44221 Germany; \\e-mail: amitava@cubmb.ernet.in}
 and Monoranjan Guchhait\\}
\vspace*{3mm}
Physics Dept., Jadavpur University, Calcutta - 700 032, INDIA \\
\vspace*{5mm}
{\large Manuel Drees\footnote{Heisenberg Fellow}}\\
\vspace*{3mm}
Dept. of Physics, Univ. of Wisconsin, Madison, WI 53706, U.S.A\\
\vspace*{1cm}
\end{center}
\begin{abstract}
Relatively light sneutrinos, which are experimentally allowed, may
significantly affect the currently popular search strategies for
supersymmetric particles by decaying dominantly into an invisible channel. In
certain cases the second lightest neutralino may also decay invisibly leading
to two extra carriers of missing energy -- in addition to the lightest
supersymmetric particle (LSP) \zi\ -- the virtual LSPs (VLSPs). It is shown
that these VLSPs are allowed in supergravity models with common scalar and
gaugino masses at the unification scale for a sizable region of parameter
space and are consistent with all constraints derived so far from SUSY
searches. The pair production of right handed sleptons, which can very well be
the lightest charged SUSY particles in this scenario, at LEP 200 and their
decay signatures are discussed. The signal survives kinematical cuts required
to remove the standard model background. Charginos are also pair produced
copiously if kinematically accessible; they also decay dominantly into
hadronically quiet di--lepton + \etmiss\ modes leading to interesting unlike
sign dilepton events which are again easily separable from the Standard Model
backgrounds at LEP 200 energies.
\end{abstract}
\newpage
\pagestyle{plain}
\setcounter{page}{1}
\section*{I. Introduction}

Supersymmetry (SUSY), the symmetry which relates bosons and fermions, is
theoretically an elegant concept \cite{1}. In addition it can solve the
notorious fine--tuning problem which haunts the standard model (SM) \cite{1},
if the supersymmetric partners of the known particles have masses of the order
of one TeV or less. The lower end of the interesting mass spectrum has already
been probed at some of the presently operating accelerators like the Fermilab
Tevatron or LEP--I at CERN. Planned accelerators like LEP--II or the LHC at
CERN can probe even higher mass scales. The search for supersymmetry (SUSY) is
therefore a high priority programme.

In most searches for SUSY particles it is assumed that there is a single,
stable, weakly interacting particle, the so-called lightest supersymmetric
particle (LSP). This particle, if produced, easily escapes detection. It is
further assumed that by virtue of a conserved quantum number ($R-$parity), all
other superparticles eventually decay into the LSP. The LSP, therefore,
carries missing transverse energy \etmiss\ which is traditionally regarded as
the most distinctive signature of SUSY particles.

The minimal supersymmetric extension of the standard model (MSSM) contains
four new spin 1/2 neutral particles. They are the super--partners of the
photon, the Z boson and the two neutral Higgs bosons. Linear combinations of
these four states, the four neutralinos ($\widetilde{Z}_i$, i=1--4),  are the
physical states. In the currently favoured models, the lightest among them
(\zi) is assumed to be the LSP \cite{1}.

Recently it has been emphasised \cite{2,3} that there may exist SUSY particles
which, though unstable, {\bf decay dominantly into invisible channels}. This
occurs if the sneutrinos ($\widetilde{\nu}$) (the super--partners of the
neutrinos), though heavier than the LSP, are lighter than the lighter chargino
(\wi) or the second lightest neutralino (\zin) and are much lighter than all
other SUSY particles. As a consequence, the invisible two--body decay mode
$\widetilde{\nu} \rightarrow \nu \zi$ opens up and completely dominates over
others, being the only kinematically allowed two--body decay of the
sneutrinos. The other necessary condition for this scheme to work is that the
\zi\ has a substantial Zino component. This, however, is almost always the
case as long as the gluino (the super--partner of the gluon) has a mass above
the lower bound obtained by the SUSY searches at the Tevatron \cite{4}. In
such cases the \zin, which also has a substantial Zino component, decays
primarily through the process $\zin \rightarrow \nu \widetilde{\nu}$. These
particles decaying primarily into invisible channels, hereafter called {\bf
virtual LSP's} (VLSP's), may act as additional sources of \etmiss\ and can
significantly affect the strategy for SUSY searches \cite{2}.

Another important consequence of relatively light sneutrinos is that \wi\
decays leptonically with a branching ratio (BR) $\approx 1$. This occurs via
the mode $\wi \rightarrow l \widetilde{\nu},l=e,\mu$ or $\tau$, which is the
only kinematically allowed two--body decay. This is to be contrasted with the
conventional scenario where chargino branching ratios closely follow those of
the $W$ bosons, in which case a mixed (leptons $+$ jets $+\etmiss$) final
state offers the best signal for chargino pair production at $e^+ e^-$
colliders \cite{tatrep}.

In ref.\cite{2} sparticle masses were treated as free phenomenological
parameters, although it was commented briefly that it is not unlikely that the
VLSP scenario can be accomodated in the N=1 SUGRA models with common squark
and gaugino masses at the GUT scale \cite{5}. In this work we show in detail
that this indeed is the case for a reasonably large region of the SUSY
parameter space, as described in section II. We also discuss in some detail
the impact of this scenario on SUSY searches at LEP 200. In this model there
are two viable candidates for the lightest charged SUSY particle: i) the
`right- handed' sleptons $\tilde{l}_R$ and ii) the lighter chargino \wi. In
this scenario both chargino and slepton pair production lead to final states
consisting of two oppositely charged leptons and missing $p_T$. In section III
the size of the signals for both the above cases is calculated, with detailed
discussion of the separation of the signal from the standard model background.
Our conclusions are spelt out in section IV. Explicit expressions for the
production and decay of chargino pairs, including polarization effects, are
listed in the Appendix.

\section*{II. The Allowed Region of Parameter Space}
We  assume a minimal N=1 supergravity model \cite{5} with a common sfermion
mass $m_0$ and a common gaugino mass $m_{1/2}$ at the GUT scale \mx. We also
assume minimal particle content. The neutralino mass matrix in the basis
$(\tilde{B}, \widetilde{W}_3, \tilde{h}_1^0, \tilde{h}_2^0)$ is then given by,
in the convention of ref.\cite{1}:
\be \label{e1}
{\cal M}^0 = \mbox{$ \left( \begin{array}{cccc}
M_1 & 0 & -M_Z \sw \cb & M_Z \sw \sib \nonumber \\
0 & M_2 & M_Z \cw \cb & -M_Z \cw \sib \nonumber \\
-M_Z \sw \cb & M_Z \cw \cb & 0 & -\mu \nonumber \\
M_Z \sw \sib & -M_Z \cw \sib & -\mu & 0 \end{array} \right). $} \ee
Here, $M_1$ and $M_2$ are SUSY breaking $U(1)$ and $SU(2)$ gaugino masses,
$\mu$ is the supersymmetric Higgs(ino) mass and $\tanb\equiv \langle H_2^0
\rangle / \langle H_1^0 \rangle$ is the ratio of the vacuum expectation values
(vevs). The assumed unification of gaugino masses leads to the following
relation at the weak scale:
\be \label{e2}
M_1 = \frac {5}{3} \tan^2 \! \theta_W M_2 \simeq 0.5 M_2, \ee
where \be \label{e3}
M_2(M_Z) = 0.82 m_{1/2} \ee
gives the connection to the GUT scale gaugino mass. The same parameters
that enter eq.(\ref{e1}) also determine the chargino masses.

The relevant slepton masses \cite{5} at the weak scale are determined by $M_2,
\ m_0$ and \tanb:
\ben \label{e4} \beq
m^2_{\tilde{l}_R} &= m_0^2 + 0.223 M_2^2 + \sin^2 \theta_W D_Z, \label{e4a} \\
m^2_{\tilde{l}_L} &= m_0^2 + 0.773 M_2^2 + (0.5-\sin^2 \theta_W) D_Z,
\label{e4b} \\
m^2_{\tilde{\nu}} &= m_0^2 + 0.773 M_2^2 - 0.5 D_Z, \label{e4c} \eeq \een
where \be \label{e5}
D_Z = M_Z^2 \frac { \tan^2 \beta - 1 } { \tan^2 \beta + 1} > 0 \ee
for $\tanb > 1$. Notice that $\msll > \mslr$ always.

Our free parameters are thus $m_0, \ M_2$ (which we traded for $m_{1/2}$),
$\mu$ and \tanb.\footnote{We do not extend the assumption of degenerate scalar
masses at the GUT scale into the Higgs sector, since Higgs bosons play no role
in our analysis; hence we cannot use the mechanism of radiative symmetry
breaking to determine the parameter $\mu$ for given SUSY breaking parameters
and top mass. We note, however, that the minimal boundary condition for Higgs
masses at the GUT scale, $m^2_H = m_0^2 + \mu^2$, does allow for VLSP
scenarios.} There are two sets of constraints on the allowed parameter space:
direct experimental constraints (primarily from LEP--I \cite {6}), and
requirements for having VLSP's. We took the following experimental constraints
into account \cite{6}:
\beq \label{e6}
&\mslr > 45 \ {\rm GeV}, \ \ \msn > 40 \ {\rm GeV}, \ \ \mwi > 46 \ {\rm GeV},
\nonumber \\
&\Gamma(Z \rightarrow \zi \zi ) < 12 \ {\rm MeV}, \ \
\sum_{i,j} \Gamma(Z \rightarrow \widetilde{Z}_i \widetilde{Z}_j) < 0.25 \
{\rm MeV},
\eeq
where the sum does not include $(i,j) = (1,1)$. The exact bounds on the
invisible and visible $Z$ decay widths vary a bit with time, but this does not
change the excluded region significantly. It was noted in ref.\cite{2} that in
the presence of VLSPs the current lower bound on the gluino mass
$m_{\tilde{g}}$ \cite{4} is likely to be relaxed since the lighter chargino
arising from gluino production decays primarily into (soft) leptons rather
than jets. We, therefore, do not take this bound at its face value. Assuming
the gluino mass to be unified with the other gaugino masses, a bound
$m_{\tilde{g}}> 120$ GeV would correspond to something like $M_2 > 35$ to 40
GeV, depending on $\alpha_S$; our explicit examples discussed in sec.III
respect this lower bound.

The defining property of the VLSP scenario is that both the sneutrino and
\zin\ decay invisibly. This implies the following constraints:
\beq \label{e7}
&\msn < \mzin < \msll, \ \mslr \nonumber \\
&\msn < \mwi < \msll.
\eeq
The second constraint in (\ref{e7}) is included in order to make \wi\ decay
via the two body mode discussed above. It does not involve \mslr\ since \slr\
does not couple to \wi.

Notice that (\ref{e7}) implies $\msn < \mslr$; eqs.(\ref{e4}) then give an
{\it upper bound} on $M_2$:
\be \label{e8}
|M_2| \leq 1.15 \sqrt{D_Z}.
\ee
This in turn requires $M_2$ to be rather small, very likely yielding a gluino
within the striking range of the Tevatron. For example, with $\tanb=2$ (10)
the upper bound on the gluino mass is about 245 (285) GeV. For each value of
$M_2$ there is an upper bound on the lighter chargino mass; the above bounds
imply $\mwi \leq 95$ (102) GeV for $\tanb=2$ (10). Then (\ref{e7}) implies
that $m_0$ cannot be too large, either, hence sleptons will also be light. The
lightest charged sparticle mass is then expected to be around $M_Z$. The
masses are, therefore, in the region of interest for LEP--II.

Figures 1a,b,c, show the allowed region in more detail. Here we fixed \tanb,
and plotted the allowed region in the $(\mu, M_2)$ plane for various values of
$m_0$. The dotted curve delineates the region excluded by sparticle searches
at LEP--I. The allowed region for fixed $m_0$ is not very large, although the
fraction of the plane with a VLSP scenario for {\it some} value of $m_0$ is
substantial, given only that $M_2$ satisfies (\ref{e8}). The results are
summarised below.

For $\tanb = 2, \mu > 0$, the left boundary of the allowed region comes from
the requirement $\msn < \mwi$; the upper boundary comes from $\mslr > \mzin$;
and the lower boundary comes from $\msn > 40$ GeV. Of course, the LEP searches
also take a bite out of the parameter space, as indicated.

For $\tanb = 2, \mu < 0$: The almost vertical left boundary for small $m_0$
comes from $\mzin < \mslr$; the top--right boundaries come from $\msn <
\mzin$; and the lower boundaries usually come from $\msn > 40$ GeV. The
situation for $\tanb = 10$ is similar.

Almost any value of $\mu$ can accomodate VLSPs if $M_2$ and $m_0$ are chosen
properly. In fig.2 we have plotted the allowed region in the $(m_0, M_2)$
plane after scanning over all $\mu$; in other words, the curves enclose the
region where a value of $\mu$ can be found so that a valid VLSP scenario
emerges. The upper bound on $M_2$ is basically just given by (\ref{e8}), i.e.
this bound can be saturated. However, for a given $m_0$, there's also a {\em
lower} bound on $M_2$: If $m_0$ is small one needs a sizable $M_2$ to get the
sneutrino to be sufficiently heavy (the negative $D-$term has to be
compensated). For larger $m_0$, one needs $M_2$ sufficiently large so that
$\mwi, \mzin > \msn$; indeed, this gets into conflict with (\ref{e8}) for $m_0
> 80$ GeV.

We also searched for the maximal and minimal allowed values, within the VLSP
scenario, of certain (differences of) sparticle masses. The lightest charged
sparticle turns out to be either a \slr\ or a \wi, if we ignore the
possibility of a light stop. The bounds on their masses depend on \tanb, due
to the constraint (\ref{e8}). We find 49 GeV $\leq m_{\tilde{e}_R} \leq$ 95
GeV for $\tanb=2$, and 59 GeV $\leq m_{\tilde{e}_R} \leq$ 103 GeV for
$\tanb=10$. We have already given the upper bounds on \mwi; we find that,
unlike for sleptons, the experimental lower bound of 46 GeV can always be
saturated for charginos. We conclude that LEP--II will only be able to probe
the entire VLSP region if the centre--of--mass energy is raised substantially
above the currently foreseen value of 176 GeV. The bounds on the masses of
left--handed sleptons lie about 20 to 30 GeV above those for
$m_{\tilde{e}_R}$.

For small \tanb, the $\slr-\zi$ mass difference can be quite small, down to
about 10 GeV for $\tanb=2$, but for $\tanb=10$ it amounts to at least 23 GeV;
this has immediate bearing on the spectrum of leptons in slepton pair events.
In contrast, the bounds on the $\wi - \tilde{\nu}$ mass difference are almost
independent of \tanb; the lower bound is always zero, indicating that leptons
produced in \wi\ decays can be very soft, while the upper bound is around 30
GeV in the VLSP scenario. Finally, while the upper bounds on $m_{\tilde{e}_R}$
and \mwi\ are very similar, we find that substantial mass splittings between
these states are possible; for $\tanb=2$, right--handed sleptons could lie
more than 20 GeV below or more than 25 GeV above the light charginos, while
for large \tanb, right--handed sleptons are almost always heavier, by as much
as 37 GeV. In order to cover the entire parameter space one therefore has to
search for both \slr\ and \wi\ pair production, which we discuss in the next
section.

\section*{III. The Di--lepton Cross--section}
\subsection*{a) Slepton pair production}
The differential cross--section $d \sigma (\eplem \rightarrow \slrp \slrm) /
dt$ is well--known \cite{7}. Of course, the cross--sections for selectrons and
smuons differ, since in the former case one has neutralino $t-$channel
exchange diagrams which do not exist for smuons. For this reason the smuon
pair cross--section depends only on the smuon mass, while the selectron pair
cross--section also depends on neutralino masses and mixings.

This is demonstrated in figs 3, which shows the dependence of the two pair
cross--sections on $M_2$ [or equivalently on the slepton mass through
eq.(\ref{e4a})] for $\tanb=2$ and various combinations of $m_0$ and $\mu<0$.
The selectron pair cross--section is even larger for $\mu>0$, and depends only
weakly on \tanb. Both the selectron and the smuon pair cross--section depend
strongly on the mass (there is a $v^3$ factor, where $v$ is the slepton's cms
velocity), but the selectron cross--section is always larger, at least at those
rather high energies. Overall, the cross--section away from the threshold is in
the pb region.

One characteristic feature of the VLSP scenario is that the charged sleptons
can always decay into at least two types of neutralinos (\zi\ and \zin), since
$\mslr > \mzin$. Expression for the corresponding partial widths can be found
in ref.\cite{7}. The decay of \slr\ into a lepton and \zin\ can in principle
provide a nice test of this scenario. Unfortunately an explicit calculation
reveals that the corresponding branching ratio is always $\leq 1\%$, and thus
too small to be observable.

Having computed total cross--sections, we have to look at the signal for
slepton pair production. Since $BR(\slr \rightarrow \zi + l) \simeq 1$, the
existence of VLSPs has actually little effect here, although it does restrict
the parameter space as discussed in the previous section. For the purpose of
illustration we have chosen two points in parameter space where \slr\ is
indeed the lightest charged sparticle. The beam energy is chosen such that no
other charged sparticles can be produced.

The first choice is $m_0 = 40$ GeV, $M_2 = -\mu = 70$ GeV, \tanb=2, which
gives \mslr=62 GeV, \mzi=39.5 GeV and $\mzin = 58.9$ GeV. In figs. 4 a--c we
present some distributions for the final state leptons coming from \slr\
decay. In these figures, the solid histogram is for selectrons, the dashed one
for smuons, and the dotted one for staus, where in the latter case only
leptonic $\tau$ decays have been included. We have applied some cuts to get
rid of $\gamma \gamma$ and $\tau^+ \tau^-$ backgrounds: we require each final
state lepton to have at least 2 GeV transverse momentum, and require the
missing transverse momentum to exceed 5 GeV; this should reduce $\gamma \gamma
\rightarrow l^+ l^-$ backgrounds to an insignificant level. Finally, we
required the opening angle of the two leptons in the transverse plane,
$\phi_{l^+l^-}$, to be less than 160$^\circ$; this value has been chosen such
that $\eplem \rightarrow \tau^+ \tau^-$ backgrounds are removed entirely.
About 85\% of selectron and smuon pair events pass those cuts, so that the
signal is largely unaffected  for those cases. However, only 35\% of leptonic
stau pair events pass the cuts, making it quite doubtful whether stau pairs
will be observable at all in this channel.

Fig. 4a shows the energy distribution of the charged leptons in the lab frame.
Before cuts this distribution is totally flat for selectrons and smuons, since
scalars decay isotropically. The cut on the transverse opening angle tends to
remove events where both leptons are emitted in the same direction as the
sleptons are going, which gives the maximal energy for the leptons. Hence the
distributions slope downwards a little bit. It should still be quite easy to
determine the endpoints of the spectrum, however, which allows one to
determine both \mslr\ and \mzi. Notice the little solid and dashed histograms
at low energies; they come from $\slr \rightarrow \zin$ decays, the Br for
which is about 0.04\%. Unfortunately these events are totally swamped by stau
pair events, so even with almost infinite statistics one can probably not see
$\slr \rightarrow \zin$ decays in our scenario. The reason is that the lepton
spectrum for the stau pair events is quite soft. This is mostly due to the
additional neutrinos, of course, which originate from the leptonic decays of
the $\tau$ leptons. However, it is also important to include the fact that the
$\tau$ leptons are produced with right--handed polarization, which means that
the charged lepton in their decay is dominantly emitted in the direction {\em
opposite} to the $\tau$ momentum, giving a soft spectrum. Given the small size
of the leptonic stau pair cross--section (remember that for any one channel,
say \eplem, an additional factor of 1/4 has to be applied), stau pairs should
not obscure the lower edge of the lepton spectrum from selectrons and smuons,
at least.

Fig. 4b shows the missing $p_T$ spectrum. Obviously one could make this cut
harder, if necessary, without losing much signal (except staus again).
Finally, Fig. 4c shows the one distribution where selectron and smuon pair
events differ in shape as well as normalization: The distribution in $\cos \!
\theta_l$, which is the angle between the final and initial negative lepton.
For selectrons this is peaked at small angles (large $\cos \! \theta_l$), due
to the $t-$channel diagrams which favor the negative selectron to go in the
direction of the initial electron. In contrast, smuon pair production has the
typical $p-$wave angular distribution for the smuons; this is largely smeared
out by smuon decay, however, giving a rather flat distribution. The
distribution in the transverse opening angle $\phi_{l^+l^-}$ (not shown) is
also quite flat, and has almost the same shape in $\tilde{e}_R$ and
$\tilde{\mu}_R$ events.

The second set of parameters we looked at in some detail is $m_0 = 65$ GeV,
$M_2 = 80$ GeV, $\mu = -105$ GeV and \tanb=2, which gives \mslr=82.5 GeV,
\mzi=44.5 GeV, and $\mzin = 82.4$ GeV. Here the $Br(\slr \rightarrow \zin)$ is
about $10^{-9}$, so those decays happen basically never. We chose
$\sqrt{s}=180$ GeV to stay below the thresholds for other sparticle pair
production; as a result the total cross--sections are quite low here. The
primary interest in this study stems from the fact that we are now above the
\ww\ threshold, so that leptonic $W$ decays are a potentially severe
background. It however transpires that this backgound can be handled adeqately
since in $\eplem \rightarrow \ww \rightarrow l^+ l^- \nu \bar{\nu}$ events one
can reconstruct the $W$ momenta, using the invariant mass and beam energy
constraints. We have kept  only those signal events where this reconstruction
is {\em not} possible; this removes the \ww\ background completely, at least
for an ideal detector.\footnote{This analysis ignores smearing of the
effective beam energy due to initial state radiation. However, not much
radiation is expected as long as the beam energy is not much larger than the
masses of the $W$, slepton or chargino.} The cuts on $p_T$ and missing $p_T$
are increased to 3 and 6 GeV, respectively, while the cut on $\phi_{l^+l^-}$
is relaxed to $\phi_{l^+l^-} \leq 165^{\circ}$, in order to account for the
increased beam energy. The overall efficiency for selectron and smuon pairs is
still about 43\%. The leptons from stau pairs are almost always too soft to
possibly come from \ww\ events, so the efficiency for leptonic stau events is
now 41\%, almost the same as for the other sleptons.

The resulting lepton energy spectrum and angular distribution are shown in
Figs. 5a,b. We see that the cut against \ww\ backgrounds distorts the lepton
spectrum quite a bit; events with harder leptons are more likely to look like
\ww\ events. At least for selectron pairs there should still be sufficiently
many events near the upper edge of the spectrum to determine it (and hence the
masses) with good precision. The missing $p_T$ spectra (not shown) actually
became a bit harder due to the cuts. The angular distribution of the leptons
is now quite flat even for selectron events; this is again due to the cut
against \ww\ pairs, which preferrably removes events with very hard leptons.
In those events the lepton and the slepton tend to go in the same direction,
i.e. the lepton ``remembers" the direction of the slepton. For the same reason
the $\phi_{l^+l^-}$ distribution (not shown) now peaks at small values of the
opening angle. The effects of the cut against the \ww\ background depend
strongly on $\sqrt{s}, \ m_{\tilde{e}_R}$ and \mzi. In particular,
combinations of parameters with small $\slr-\zi$ mass difference, and hence
rather soft leptons, now actually have higher efficiencies and less distorted
distributions.

\subsection*{b) Chargino pair production}
The cross--section for \wi\ pair production at LEP 200 energies is typically of
the order of a few pb \cite{8}, compared to about 16 pb for \ww\ production.
The branching ratio of chargino decays in the leptonic channel in the MSSM
with heavy squarks and sleptons is roughly the same as that of the $W$'s. In
this case chargino search in the dilepton channel faces a severe background,
and the best signal can be found in the mixed channel where one chargino
decays hadronically and the other leptonically \cite{tatrep}. However, in the
VLSP scenario a chargino pair decays into a pair of stable, oppositely charged
leptons ($e$ or $\mu$) with a large branching ratio (approximately 4/9). The
corresponding branching ratio for \ww\ pairs is only about 0.05. Thus the
prospect of \wi\ search in the dilepton channel is more promising in this
scenario. (Of course, now charginos can only produce hadrons from tau lepton
decays, making the chargino signal in the mixed mode much more difficul to
find.)

Since the dilepton final state in this case arises from the decay of spin 1/2
charginos, the distributions of the final state leptons depend on the
polarizations of the charginos. These polarizations can be conveniently taken
into account by calculating the di--lepton cross--section and related
distributions using the helicity projection technique \cite{9}. The
calculations are tedious but straightforward and the results are given in the
Appendix. The spin--averaged cross--section for chargino pair production is
well--known \cite{8}. We have checked that the total number of dilepton events
resulting from our calculations agree well with that calculated from the
cross--section of ref.\cite{8} multiplied by the appropriate branching ratio
(4/9). However, polarization affects the distributions of the final state
leptons and their response to kinematical cuts significantly.

As discussed in Sec.II, the lighter chargino is one of the two candidates for
the lightest charged sparticle within the VLSP scenario. Chargino production
can be most conveniently studied by choosing the beam energy below the
threshold for the production of other charged sparticles. Di--lepton final
states consisting of any combination of $e$ and $\mu$, after removing the \ww\
and other backgrounds by the kinematical cuts discussed in the previous
subsection, then provide an unambiguous signal of chargino pair production.
For beam energies above the slepton threshold, $e -\mu$ pairs still provide a
clean signature for chargino production. Such final states cannot arise from
slepton pairs whose decays conserve flavour. In the following we shall present
two illustrative examples of chargino signals in the VLSP model.

As our first case we have chosen $m_0 =67.5$ GeV, $M_2= 53$ GeV, $\tanb= 2.0$
and $\mu = -150$ GeV, giving $m_{\tilde{\nu}} = 65.4$ GeV, $m_{\tilde{l}_R} =
79.6$ GeV, $m_{\tilde{l}_L} = 90.0$ GeV and $\mwi = 73.7$ GeV. In this case
the dilepton cross--section arising out of direct chargino decays at the c.m.
energy $\sqrt{s}=160$ GeV is 0.15 pb. The kinematical cuts used are the same
as the ones discussed in the last section (the cut against the \ww\ background
is of course irrelevant in this case). Thus for an integrated luminosity of
500 pb$^{-1}$ we expect about 75 events. Since $2 m_{\widetilde{l_R}} >
\sqrt{s}$, any combination of $e$ and $\mu$ contributes to the signal.

Since the charginos can also decay into $\tau$'s, stable leptons from the
decays  $\tau \rightarrow \ (e \;{\rm or}\; \mu)\; \nu \;\bar{\nu}$ can enhance
the signal. The cross--section for this process, however, is not significant
when both the charginos decay into $\tau$'s due to the small leptonic
branching ratio of the $\tau$ as well as the reduced efficiency for these
rather soft events. On the other hand, the contribution from events where only
one  of the produced charginos decays into a $\tau$ while the other directly
decays into $e$ or $\mu$ is not negligible. The resulting dilepton
cross-section for the choice of parameters and cuts given above is 0.02 pb.

In figure 6a we show the energy distribution of charged leptons from direct
chargino decays, for the above choice of parameters (lower solid line). The
end points of this spectrum are particularly interesting, since they determine
both $m_{\tilde{\nu}}$ and \mwi. Notice that the spectrum rises with
increasing $E_l$; this is due to the polarization of the produced charginos.
As discussed above, stable leptons arising from $\tau$ decays can in principle
obscure the lower edge of the spectrum. The distortion of the energy
distribution in the presence of a single tau decay is also plotted in the same
figure (upper solid line). It is clear that the number of the latter events is
not large enough to obscure completely the characteristics of the energy
spectrum of the leptons arising from direct chargino decays. Therefore, one
should be able to determine \mwi\ and $m_{\tilde{\nu}}$ from the energy
distribution with reasonable accuracy.

We note in passing that in the above example the mass difference between the
chargino and the sneutrino is rather small. The clean environment provided by
an $e^+e^-$ collider nevertheless allows us to derive a viable signal. This is
to be contrasted with chargino searches in this channel at the Tevatron, where
the signal can be completely washed out if the mass difference is small
\cite{10}. This is due to the fact that much stronger $p_T$ cuts on the final
state leptons are required in this case to eliminate backgrounds.

The second set of parameters we have considered in some detail is $m_0 =40$
GeV, $M_2= 70$ GeV, $\tanb = 2.0$ and $\mu = -75$ GeV. This yields the
following sparticle spectrum: $m_{\tilde{\nu}} = 53.8$ GeV, $m_{\tilde{l}_R}
= 61.9$ GeV, $m_{\tilde{l}_L} = 82.0$ GeV and $\mwi = 79.9$ GeV. This case is
interesting for two reasons. Firstly, the beam energy has to be above the \ww\
threshold and the cuts against the \ww\ background are necessary. Secondly,
since $m_{\tilde{l}_R} < \mwi$, only $e - \mu$ final states should be
considered for the chargino signal. A calculation using the standard cuts
yields a dilepton cross--section of 0.13 pb from direct decays at c.m.
energy $\sqrt{s} =180$ GeV. Thus for an integrated luminosity of 500 pb$^{-1}$
we expect about 65 events. The cross--section for the same final state arising
due to intermediate $\tau$ decays is 0.04 pb. However, we again find that such
decays cannot obscure the characteristics of the $e - \mu$ energy spectrum
(see the dashed histograms in Fig. 6a).

The angular distribution of the negative lepton relative to the electron beam
for the two cases is shown in figure 6b. Case 2 has higher beam energy and a
lighter sneutrino; this enhances the contribution of the sneutrino $t-$channel
exchange diagram, leading to a peak at $\cos \! \theta = 1$. This peak would
be even more pronounced without our cut against the \ww\ background, which
again mostly removes events with hard leptons, which dominantly go in the
direction of the parent chargino. However, by comparing Fig. 6a with Fig. 5a
we see that here the leptons are significantly softer than in the slepton
production example we discussed above, so that the distortion due to the cut
against the \ww\ background is less severe. In contrast, in case 1 the angular
distribution is peaked in the backward direction. This is due to destructive
interference between $s-$ and $t-$channel diagrams in the forward direction.
Notice that the light chargino has a significantly larger higgsino component
in case 2 than in case 1; this changes the relative importance of the various
contributions to the cross--section. Finally, the effect of events where one
chargino decays into a tau lepton which in turn decays into an $e$ or $\mu$ is
now mostly an increase of the total rate, with only minor effects on the shape
of the distribution.

We have also checked the cross--section after cuts for several representative
points of the parameter space for $\tanb = 10$. Results at $\sqrt{s} =150$ and
180 GeV are shown in table 1. Here $N_A$ and $N_B$ refer to the number of
direct and $\tau$ mediated di--lepton events, respectively, for an integrated
luminosity of 500 $pb^{-1}$. If $\mwi > m_{\tilde{l}_R}$, only $e-\mu$ events
are considered. We see that a viable signal can be expected for this value of
\tanb\ as well. We have also checked that, inspite of the $\tau$ mediated
events, the end points of the lepton energy spectrum in each case are
sufficiently well defined to allow a determination of \mwi\ and
$m_{\tilde{\nu}}$.

\section*{IV. Conclusions}
In conclusion, we reiterate that the VLSP scenario can drastically alter
search strategies for SUSY particles at future colliders, due to the presence
of additional carriers of \etmiss\ and the enhanced leptonic branching ratio
of the lighter chargino. One purpose of this paper is to show that this
scenario is consistent even with the $N=1$ SUGRA models with highly constrained
mass spectra for the SUSY particles. In this scenario either the right--handed
slepton or the lighter chargino turns out to be lightest charged SUSY
particle.

It has been known for some time that both can be pair produced with sizable
cross--sections at LEP-II, if they are not too heavy. In the VLSP scenario,
both slepton and chargino production are signalled by a pair of charged
leptons and missing $p_T$, which can be easily disentangled from the standard
model background by the kinematical cuts discussed in the text. In particular,
we suggested to remove \ww\ backgrounds by only accepting events which {\em
cannot} be interpreted as coming from $W$ pairs. In principle this removes
this background completely, leaving at least 40\% of the signal behind. We
found that the presence of a relatively light second neutralino has little
effect on the production or decay of right--handed sleptons. Light sneutrinos
play no role at all in the slepton signal, but they do have dramatic effects
on the signal for the pair production of light charginos, which now nearly
always decay into a sneutrino and a charged lepton. The cleanest signal again
comes from di--lepton final states containing only $e$ and $\mu$, where the
contribution from $\wi \rightarrow \tau \rightarrow e,\mu$ decays is also
significant. The most straightforward way to distinguish between sleptons and
charginos in this scenario is that the former always give lepton pairs of the
same flavour (up to a very small contribution from $\tilde{\tau}$ pair
production), while chargino pairs are equally likely to result in
unlike--flavoured lepton pairs. Finally, while LEP--II will almost certainly
probe a significant region of parameter space where VLSPs exist, a full
exploration is only possible if the centre--of--mass energy can be raised to
210 GeV or so.

Unfortunately the presence of a light \zin\ has no direct impact on the
chargino signal. Within the MSSM one might be able to infer its presence from
slepton production by trying to fit the parameters $M_2,\ \mu$ and \tanb\
describing the neutralino sector using the measured values of the lightest
neutralino mass, the total $\tilde{e}_R$ production cross--section, and the
angular distribution of the electrons. A careful analysis of signal and
backgrounds, including detector simulation, is necessary to decide to what
extent this is feasible.

In this paper we focussed on di--lepton final states. In the VLSP scenario we
also expect enhanced signals for events with single photons and missing
energy \cite{ddr}. Not only \zi\ pairs, but also $\zi \zin, \ \zin$ pairs, and
sneutrino pairs contribute to this signal. Unfortunately the contribution from
neutralino production turns out to be quite small compared to the SM and
sneutrino contributions; the single photon signal is therefore also not very
sensitive to the presence of a light \zin.

\subsection*{Acknowledgements:}
A.D.'s work was supported by the Department of Science and Technology,
Government of India, and the Alexander von Humboldt Stiftung, Bonn. The work of
M.G. has been supported by the Council for Scientific and Industrial Research,
India. The work of M.D. was supported in part by the U.S. Department of Energy
under grant No. DE-FG02-95ER40896, by the Wisconsin Research Committee with
funds granted by the Wisconsin Alumni Research Foundation, as well as by a
grant from the Deutsche Forschungsgemeinschaft under the Heisenberg program.
We wish to thank the organizers of the Workshop on High Energy Particle
Physics (WHEPP3), held in Madras in January 1994, where this project was
initiated.

\renewcommand{\theequation}{A.\arabic{equation}}
\setcounter{equation}{0}
\section*{Appendix}
In the following we summarise the derivation of the matrix element squared for
the process $\eplem \rightarrow \widetilde{W}^+_1 \widetilde{W}_1^-$, with
subsequent decay $\widetilde{W}_1^+ (\widetilde{W}_1^-) \rightarrow \l^+
\tilde{\nu} (l^- \tilde{\nu}^*) $, where $l = e$ or $\mu$.

The amplitude for the production of a chargino pair
$\widetilde{W}_1^+ (p_3, \lambda_3)$
and $\widetilde{W}_1^- (p_4, \lambda_4)$ with momenta and helicities as
indicated are given by (the subscripts in the following expressions
refer to the exchanged particles in the $s-$ or $t-$channel):
\begin{eqnarray}
A_{\tilde{\nu}}(\lambda_3, \lambda_4) &\equiv &
A_{1}(\lambda_3, \lambda_4) \nonumber \\
&=& C_{\tilde{\nu}} \bar u_{\widetilde{W_1}} (p_4, \lambda_4)\:P_L u_{e}
( p_2, s_2)
\bar v_{e}( p_1, s_1) \:P_R v_{\widetilde{W_1}}(p_3, \lambda_3)\\
A_{\gamma, Z }(\lambda_3, \lambda_4) &\equiv &
A_{2,3}(\lambda_3, \lambda_4) \nonumber \\
&=& C_{\gamma, Z} \bar u_{\widetilde{W_1}} (p_3, \lambda_3)\:
\Gamma^{\mu}_{\gamma, Z}
v_{\widetilde{W_1}}(p_4, \lambda_4) \: \bar v_{e}( p_1, s_1)\Gamma^{\prime}_
{\mu\:\:\gamma, Z}\: u_{e}( p_2, s_2)
  \label{ampli}
\end{eqnarray}
In the above the chargino spinors (with subscript \wi) are helicity spinors.
Spin averaging will be done in the matrix element squared for the initial
\eplem\ pair (denoted by the subscript $e$). We have used the following
abbreviations:
\begin{eqnarray}
C_{\widetilde{\nu}}& =& \frac{-i g^2 |V_{11}|^2 }
{t - m_{\widetilde{\nu}}^2}\\
C_{Z}& =& \frac {i g^2 } {c_w^2 (s - m_{Z}^2)}\\
C_{\gamma}& =& \frac {i e^2 } {s}\\
\Gamma^{\mu}_{\gamma}&=& \Gamma^{\mu \prime}_{\gamma}= \gamma^{\mu}\\
\Gamma^{\mu}_{Z}&=& \gamma^{\mu} (O^{L}_{11} P_{L} +O^{R}_{11} P_{R})\\
\Gamma^{\mu \prime}_{Z}&=& \gamma^{\mu} ( c_v + c_a \gamma_5)
\end{eqnarray}
where $g$ is the coupling constant of $SU(2)_L$, $s_w (c_w)$ is the sine
(cosine) of the weak mixing angle, $s=(p_1 + p_2)^2$, $t=(p_1 - p_3)^2$, $c_v
= -0.25+ s_w^2$, and $c_a = 0.25$. The factors $|V_{11}|$ and $O^{L,R}_{11}$
enter in the $\wi - l - \tilde{\nu}$ and $\wi - \wi - Z$ couplings,
respectively \cite{1,8}. They are obtained from the diagonalisation of the 2 x
2 chargino mass matrix. We have followed the notation and prescription of
\cite{1}.

In this formalism two--body decay amplitudes of positive and negative
charginos are given by
\begin{eqnarray}
D_+(\lambda_3)&=& -i g V_{11}^{*} \bar u_{l}( p_5, s_5) \:P_L
u_{\widetilde{W_1}}(p_3, \lambda_3)\\
D_-(\lambda_4)&=& -i g V_{11} \bar v_{\widetilde{W_1}}(p_4, \lambda_4) \:P_R
v_{l}( p_7, s_7)
\end{eqnarray}
In the  matrix element squared the final state lepton (denoted by the
subscript $l$) spins will be summed.

The full matrix element squared for a particular helicity configuration of the
charginos is of the form [after averaging (summing) over the initial (final)
lepton spins]:
\begin{eqnarray}
M_{i j} (\lambda_3,\lambda_3^{\prime}; \lambda_4 , \lambda_4^{\prime})&=&
{1 \over 4} \sum_{s_{1}, s_{2}}\:\:\sum_{s_{5}, s_{7}} A_{i} (\lambda_{3}
, \lambda_{4}) \: D_{+}(\lambda_{3}) D_{-}(\lambda_{4})\nonumber\\
& & \times A_{j}^{*} (\lambda_{3}^{\prime}
, \lambda_{4}^{\prime}) \: D_{+}^{*}(\lambda_{3}^{\prime}) D_{-}^{*}
(\lambda_{4}^{\prime})
\label{summed}
\end{eqnarray}
We have not included the Breit--Wigner propagators of the two charginos. It is
understood that after using the narrow width approximation on them the total
phase space for the $2 \rightarrow 4$ process can be factorised into a product
of three phase spaces: a two--body phase space for the production of the
charginos and two two--body phase spaces for their decays.

Using standard manipulations we obtain the following matrix element squared
for the production of chargino pairs:
\begin{eqnarray}
 \frac {1} {4}   \sum_{s_{1}, s_{2}}
|A_{1}|^2(\lambda_3,\lambda_3^{\prime};\lambda_4, \lambda_4^{\prime})
& = &
\frac {1} {4}\: |C_{\tilde{\nu}}|^2 {\rm Tr} \left[
v_{\widetilde{W_1}}(p_3, \lambda_3)
\bar{v}_{\widetilde{W_1}}(p_3, \lambda_3^{\prime})\:
P_L \not{\!p_1} \right] \nonumber\\
&\cdot & {\rm Tr}
\left[ u_{\widetilde{W_1}}(p_4, \lambda_4^{\prime}) \bar{u}_{\widetilde{W_1}}
(p_4, \lambda_4)
P_{L} \not{ \! p_{2}} \right] \\
 \frac {1} {4}   \sum_{s_{1}, s_{2}}
|A_{2, 3}|^2(\lambda_3, \lambda_3^{\prime};
\lambda_4,  \lambda_4^{\prime}) & = &
\frac {1} {4}\: |C_{\gamma,Z}|^2 {\rm Tr}
\left[ \not{ \! p_{1}} \Gamma^{\prime}_{\mu\:\:\gamma, Z}
 \not{ \! p_{2}} \Gamma^{\prime}_{\nu\:\:\gamma, Z} \right]  \nonumber \\
&\cdot &{\rm Tr} \left[ u_{\widetilde{W_1}}(p_3,
\lambda_3^{\prime})
\bar{u}_{\widetilde{W_1}}(p_3, \lambda_3)\: \Gamma_{\mu\:\:\gamma, Z}
\nonumber \right. \\ &&{} \left.  \hspace*{6mm} \cdot
v_{\widetilde{W_1}}(p_4, \lambda_4) \bar{v}_{\widetilde{W_1}}(p_4,
\lambda_4^{\prime})
\Gamma_{\nu\:\:\gamma, Z} \right]  \\
 \frac {1} {4}   \sum_{s_{1}, s_{2}}
A_{1} A_{2,3}^{*}(\lambda_3, \lambda_3^{\prime};
\lambda_4 \lambda_4^{\prime}) +CT &=&
\frac {1} {4}\: C_{\widetilde{\nu}} C_{\gamma,Z}^{*}
{\rm Tr}
\left[ v_{\widetilde{W_1}}(p_4, \lambda_4) \bar{v}_{\widetilde{W_1}}(p_4,
\lambda_4^{\prime})
\Gamma_{\mu\:\:\gamma, Z} \nonumber \right. \\ &&{} \left.  \hspace*{2.1cm}
\cdot u_{\widetilde{W_1}}(p_3, \lambda_3^{\prime})
\bar{u}_{\widetilde{W_1}}(p_3, \lambda_3)\: P_R \not{ \! p_{1}}
\Gamma^{\prime \prime}_{\mu\:\:\gamma, Z} \not{ \! p_{2}} P_L \right]
\nonumber \\ \end{eqnarray}
where $\Gamma^{\prime \prime}_{\mu\:\: Z}
= \Gamma^{\prime }_{\mu\:\: Z}(c_a \rightarrow - c_a)$ ,
$\Gamma^{\prime \prime}_{\mu\:\: \gamma}
= \Gamma^{\prime }_{\mu\:\: \gamma}$
and CT stands for the conjugate term.
\begin{eqnarray}
 \frac {1} {4}   \sum_{s_{1}, s_{2}}
A_{3} A_{2}^{*}(\lambda_3, \lambda_3^{\prime};
\lambda_4 \lambda_4^{\prime}) +CT &=&
\frac {1} {4}\: C_{Z} C_{\gamma}^{*} {\rm Tr}
\left[ \not{ \! p_{1}} \Gamma^{\prime}_{\mu\:\: Z}
 \not{ \! p_{2}} \gamma_{\nu} \right]  \nonumber \\
&\cdot&{\rm Tr}
\left[ u_{\widetilde{W_1}}(p_3, \lambda_3^{\prime})
\bar{u}_{\widetilde{W_1}}(p_3, \lambda_3)\: \Gamma_{\mu\:\: Z}
v_{\widetilde{W_1}}(p_4, \lambda_4) \bar{v}_{\widetilde{W_1}}(p_4,
\lambda_4^{\prime})
\gamma_{\nu} \right] \nonumber \\
\end{eqnarray}

Similarly the M.E squared for the decays are
\begin{eqnarray}
\sum_{s_{5}} D_+(\lambda_3) D_{+}^{*} (\lambda_3^{\prime}) &=&
g^{2}\: |V_{11}|^2 {\rm Tr} \left[ v_{\widetilde{W_1}}(p_3,
\lambda_3^{\prime})
\bar{v}_{\widetilde{W_1}}(p_3, \lambda_3)\: P_L \not{\!p_5} \right]
\nonumber \\
&=& g^{2}\: |V_{11}|^2 {\rm Tr}
\left[ u_{\widetilde{W_1}}(p_3, \lambda_3) \bar{u}_{\widetilde{W_1}}(p_3,
\lambda_3^{\prime})
P_{R} \not{ \! p_{5}} \right] \\
\sum_{s_{7}} D_-(\lambda_4) D_{-}^{*} (\lambda_4^{\prime}) &=&
g^{2}\: |V_{11}|^2 {\rm Tr} \left[ v_{\widetilde{W_1}}(p_4,
\lambda_4^{\prime})
\bar{v}_{\widetilde{W_1}}(p_4, \lambda_4)\: P_R \not{\!p_7} \right]
\nonumber \\
&=& g^{2}\: |V_{11}|^2 {\rm Tr}
\left[ u_{\widetilde{W_1}}(p_4, \lambda_4) \bar{u}_{\widetilde{W_1}}(p_4,
\lambda_4^{\prime})
P_{L} \not{ \! p_{7}}\right]
\end{eqnarray}
The sum over different helicity configurations can be carried out by the
following strategy. It is sufficient to compute each of the above traces for
the configuration $\lambda_3 = \lambda_3^{\prime} = \lambda_4 =
\lambda_4^{\prime}= +$. The remaining traces can then be obtained by
inspection. Using the outer products of helicity spinors given in
ref.\cite{10}, Appendix C, p 573, a typical trace in eqs (A12) -- (A15)
reduces to the form
\begin{equation}
T= A + S_{3} \cdot X_{3} + S_{4} \cdot X_{4}+ S_{3} \cdot Y(S_{4})
\end{equation}
where $S_{3}, S_{4}$ are the covariant spin vectors of the charginos. In
eq.(A.18), $A$ is a scalar, the four vectors $X_{3}$ and $X_{4}$ are
independent of the covariant spin vectors and all terms in the four vector $Y$
are linear in $S_{4}$. Multiplying the traces by the decay terms for this
helicity configuration we obtain the following M.E. squared:
\begin{equation}
|M|^{2}= T  \left[ p_{5} \cdot (p_{3} - m S_{3})\right]
\left[ p_{7} \cdot (p_{4} + m S_{4}) \right],
\end{equation}
where $m$ is the chargino mass.

The summation over the helicities can now be carried out using the following
observation. For $\lambda_{3} = + , \lambda_{3}^{\prime} = - \ ( \lambda_{3} =
- , \lambda_{3}^{\prime}= +)$ and fixed $\lambda_{4} , \lambda_{4}^{\prime}$,
the result is obtained from the $S_{3}$ dependent terms of eqs.(A18) and (A19)
by the substitution $S_{3} \rightarrow C_{3} ( C_{3}^{*})$ where the four
vector $C$ is defined in ref.\cite{10}. For $\lambda_{3}= \lambda_{3}^{\prime}
= -$, the corresponding result can be obtained from eqs.(A18) and (A19) by the
substitution $S_{3} \rightarrow -S_{3}$. The summation over $\lambda_{3},
\lambda_{3} ^{\prime}$ can now be performed by using the identity
\begin{equation}
2 (X \cdot S_{i}) (Y \cdot S_{i})+ (X \cdot C_{i}) (Y \cdot C_{i}^{*}) +
(X \cdot C_{i}^{*}) (Y \cdot C_{i}) = (2 / m^2) (X \cdot p_{i})
( Y \cdot p_{i}) - 2 (X \cdot Y)
\end{equation}
where $X$ and $Y$ are two arbitrary four vectors and $p_{i}$  refers to the
momentum  of the chargino. After this helicity summation  we obtain from
eq.(A19):
\begin{equation}
\sum_{\lambda_{3}, \lambda_{3}^{\prime}} |M|^{2} = \left\{ \rule{0mm}{5mm}
2 (A + S_{4} \cdot X_{4})
(p_{3} \cdot p_{5}) - 2 m \left[ f(X_{3}) + f\left(Y(S_{4})\right) \right]
\right\} \left[ p_7 \cdot (p_4 + m S_4) \right],
\end{equation}

where
\begin{equation}
f(X) = \left[ (X \cdot p_3)(p_3 \cdot p_5 )/ m^2 \right] - X \cdot p_5.
\end{equation}
The summaton over $\lambda_{4}, \lambda_{4}^{\prime}$ can now be carried
out by making substitutions similar to the ones stated above
and using eq.(A20). The final result is:
\begin{equation}
\sum_{\lambda_{3}, \lambda_{3}^{\prime},\lambda_{4}, \lambda_{4}^{\prime} }
 |M|^{2} = 4 \left[ A\; p_{35}\; p_{47} - m f(X_3)\; p_{47}
-  p_{35} \: g(D_3) + m^2  g( D_5) + \;p_{35}\; m \;g( X_4) \right]
\end{equation}
where
\begin{equation}
g(X) = \frac {(X \cdot p_4) p_{47}} {m^2} - X \; \cdot \; p_7;
\end{equation}
\begin{equation}
p_{ij} = p_i \; \cdot p_j,
\end{equation}
and the four vector $D_i$ is defined by
\begin{equation}
 p_i \cdot Y(S_4) = S_4 \cdot D_i
\end{equation}
It is therefore sufficient to calculate $A$, $X_3, X_4$ and $Y(S_4)$ for each
of the terms in eqs.(A12 - 15). The computation of the trace and the
following simplifications can be easily done by using MATHEMATICA.

For the $|A_1|^2$ term (A12) the calculation is simple and one obtains
directly from eqs.(A12), (A16) and (A17) the following matrix element squared:
\begin{equation}
|M_1|^2 = \frac {1} {4} |C_{\widetilde{\nu}}|^2 (4 p_{24}\; p_{47}
- 2 m^2 \;p_{27})\; (4 p_{13}\; p_{35} - 2 m^2 \;p_{15})
\end{equation}

The relevant terms for the $s-$channel $Z$ exchange diagram are:
\begin{eqnarray}
A &=& 4 O_L^2 \left[ ( c_a+c_v)^2 p_{14} p_{23} + ( c_a-c_v)^2 p_{13} p_{24}
\right] \nonumber \\
&+& 4 O_R^2 \left[ ( c_a+c_v)^2 p_{13} p_{24}
+ ( c_a-c_v)^2 p_{14} p_{23} \right] \nonumber \\
&+& 8 m^2  O_L O_R ( c_a^2+c_v^2) p_{12} \\
X_3 &=& 4m \left\{- O_L^2  \left[ ( c_a+c_v)^2 p_{14} p_{2} + ( c_a-c_v)^2
p_{1} p_{24} \right] \nonumber \right. \\
&&{} \left. \hspace*{5mm}
+ O_R^2 \left[ ( c_a+c_v)^2 p_{1} p_{24} + ( c_a-c_v)^2
p_{14} p_{2} \right] \right\} \nonumber \\
&+& 16 m   O_L O_R  c_a c_v ( p_{1} p_{23} - p_2 p_{13}) \\
X_4 &=& 4m \left\{ O_L^2 \left[ ( c_a+c_v)^2 p_{1} p_{23} + ( c_a-c_v)^2 p_{2}
p_{13} \right] \nonumber \right. \\ &&{} \left. \hspace*{5mm}
- O_R^2 \left[ ( c_a+c_v)^2 p_{2} p_{13} + ( c_a-c_v)^2 p_{1}
p_{23}\right] \right\} \nonumber \\
&+& 16 m  O_L O_R  c_a c_v ( p_{1} p_{24} - p_2 p_{14})\\
Y(S_4) &=& -4 O_L^2 m^2  \left[( c_a+c_v)^2 (p_{1} \cdot S_4) p_{2}
+ ( c_a-c_v)^2 (p_{2} \cdot S_4) p_{1} \right] \nonumber \\
&-& 4 O_R^2 m^2 \left[( c_a+c_v)^2 (p_{2} \cdot S_4) p_{1}
+ ( c_a-c_v)^2 (p_{1} \cdot S_4) p_{2} \right] \nonumber \\
&+& 8  O_L O_R ( c_a^2 +  c_v^2) \left[ (p_{1} \cdot S_4)
(p_4 p_{23} - p_2 p_{34})
+ (p_{2} \cdot S_4) (p_4 p_{13} - p_1 p_{34} ) \nonumber \right.  \\
&&{}\left. \hspace*{2.7cm} + (p_{3} \cdot S_4)
(p_1 p_{24} + p_2 p_{14} -p_4 P_{12}) + S_4 (p_{12} p_{34} - p_{13} p_{24}
- p_{14} p_{23}) \right] \nonumber \\
\end{eqnarray}

For the $s-$channel $\gamma$ exchange diagram the corresponding terms are:
\begin{eqnarray}
A &=& 8 ( m^2 p_{12} + p_{14} p_{23} +  p_{13} p_{24})\\
Y(S_4) &=& 8 \left[\left\{ (p_3 \cdot S_4) p_{24}- (m^2 + p_{34})
 (p_2 \cdot S_4) \right\} p_1
+ (p_1 \leftrightarrow p_2) \right] \nonumber \\
&+& 8 \left[ (p_1 \cdot S_4 p_{23}
+ p_2 \cdot S_4 p_{13} - p_3 \cdot S_4 p_{12}) p_4 + ( p_{12} p_{34}
- p_{14} p_{23} - p_{13} p_{24}) S_4 \right] \nonumber \\
\end{eqnarray}

For the $\widetilde{\nu} - Z$ interference the corresponding terms are:
\begin{eqnarray}
A &=& ( c_v - c_a)( 2 O_L  p_{13} p_{24} + O_R  p_{12} m^2)\\
X_3 &=&-(c_v - c_a) m  \left[ 2 O_L  p_{24} p_{1} + O_R ( p_{1} p_{23}
- p_2 p_{13}) \right] \\
X_4 &=& (c_v - c_a) m \left[ 2 O_L  p_{13} p_{2} - O_R ( p_{1} p_{24}
- p_2 p_{14}) \right] \\
Y(S_4) &=& ( c_v - c_a) \left\{ - 2 O_L m^2  (p_{2} \cdot S_4) p_{1}
\nonumber \right.  \\
&&{}\left. \hspace*{1.5cm} +
O_R \left[ - p_2 p_{34} (p_{1} \cdot S_4) - p_1 p_{34} (p_{2} \cdot S_4)
+ (p_{3} \cdot S_4) (p_1 p_{24} + p_2 p_{14})
\nonumber \right. \right. \\ && \left. \left.
{} \hspace*{2.4cm} + \left( (p_{1} \cdot S_4) p_{23}
+ (p_{2} \cdot S_4) p_{13} - (p_{3} \cdot S_4) p_{12} \right) p_4
\nonumber \right. \right. \\ && \left. \left. {} \hspace*{2.4cm}
+ ( p_{12} p_{34} - p_{13} p_{24} - p_{14} p_{23}) S_4 \right]
\rule{0mm}{5mm} \right\}
\end{eqnarray}
%\newpage

For the $\tilde{\nu} - \gamma$ interference the corresponding terms are
\begin{eqnarray}
A &=& m^2 p_{12} + 2 p_{13} p_{24}\\
X_3 &=& - m p_1(p_{23} + 2 p_{24}) + m p_2 p_{13}\\
X_4 &=& -m p_1 p_{24} + m p_2 (2 p_{13} + p_{14})\\
Y(S_4) &=& - 2 m^2 p_1 (p_2 \cdot S_4)  - p_2 p_{34} (p_1 \cdot S_4)
- p_1 p_{34} (p_2 \cdot S_4) + p_1 p_{24} (p_3 \cdot S_4) \nonumber \\
&+&p_2 p_{14}(p_3 \cdot S_4) + p_4 p_{23}  (p_1 \cdot S_4) \nonumber \\
&+& p_4 p_{13} (p_2 \cdot S_4) - p_4 p_{12} (p_3 \cdot S_4) - S_4 p_{14} p_{23}
- S_4 p_{13} p_{24} + S_4 p_{12} p_{34}
\end{eqnarray}

For the $\gamma - Z$ interference the corresponding terms are
\begin{eqnarray}
A &=& 4 c_v ( O_L + O_R) (m^2 p_{12} + p_{14} p_{23}
+ p_{13} p_{24}) \nonumber \\
&+& 4 c_a (O_L - O_R) ( p_{14} p_{23} - p_{13} p_{24})\\
X_3 &=& - 4 c_v m (O_L - O_R)( p_1  p_{24} + p_2 p_{14}) \nonumber \\
&+& 4 c_a m (O_L + O_R) \left[ p_1 (p_{24} + p_{23})
- p_2 (p_{14}+p_{13}) \right] \\
X_4 &=&  4 c_v m (O_L - O_R)( p_1  p_{23} + p_2 p_{13}) \nonumber \\
&+& 4 c_a m (O_L + O_R) \left[ p_1 (p_{24} + p_{23})
- p_2 (p_{14}+ p_{13}) \right] \\
Y(S_4) &=& 4 c_v (O_L + O_R) \left\{ - m^2 \left[ p_2 (p_1 \cdot S_4) + p_1
(p_2 \cdot S_4) \right]  - p_{34} \left[ p_2 (p_1 \cdot S_4)
+ p_1 (p_2 \cdot S_4) \right] \nonumber \right.  \\
&&{} \left. \hspace*{2cm}+ (p_3 \cdot S_4) (p_1 p_{24} + p_2 p_{14})
+ S_4( p_{12} p_{34} - p_{13} p_{24} - p_{14} p_{23})
 \nonumber \right. \\
&&{} \left. \hspace*{2cm}
+ p_4 \left[ p_{23} (p_1 \cdot S_4) + p_{13} (p_2 \cdot S_4) -
p_{12} (p_3 \cdot  S_4) \right] \rule{0mm}{5mm} \right\} \nonumber \\
&-& 4 c_a m^2 ( O_L - O_R) \left[ p_2 ( p_1 \cdot S_4) -
p_1 (p_2 \cdot S_4) \right].
\end{eqnarray}

\clearpage
{\bf Table 1:} Number of di--lepton events from chargino pair production for
$\tanb=10$, after the cuts discussed in Sec.III have been imposed, assuming an
integrated luminosity of 500 pb$^{-1}$. $N_A$ includes only direct $\wi
\rightarrow e, \mu$ decays, while $N_B$ is the number of additional events due
to $\wi \rightarrow \tau \rightarrow e,\mu$ decays. All masses are in GeV.

\vspace*{5mm}
\begin{center}
\begin{tabular}{||c|c|c|c|c|c|c|c|c||}
\hline \hline

$m_0$ & $M_2$ &$\mu$ &$\sqrt{s}$ & \mwi\ &
$m_{\tilde{\nu}}$ & $m_{\tilde{l}_R}$ & $ N_A $ & $ N_B $\\
\hline \hline
20 & 90 &-120 & 150 &68.85 &50.93 &63.85 &41  &10 \\
20 & 90 &-120 & 180 &68.85 &50.93 &63.85 &100 &25 \\
40 & 80 &-120 & 150 &62.97 &49.8  &69.98 &90  &22 \\
60 & 80 &-200 & 150 &74.45 &66.93 &83.05 &19  &3  \\
60 & 80 &-200 & 180 &74.45 &66.93 &83.05 &55  &10  \\
20 & 88 & 140 & 150 &72.59 &48.15 &63.23 &131 &27 \\
40 & 80 & 200 & 150 &62.32 &49.8  &69.98 &133 &32 \\
60 & 80 & 400 & 150 &73.57 &66.93 &83.05 &50  &8  \\
60 & 80 & 400 & 180 &73.57 &66.93 &83.05 &125 &20  \\
\hline \hline
\end{tabular}
\end{center}
\end{document}